\documentclass[a4paper,11pt]{article}
\pdfoutput=1
            
\usepackage{jcappub}
\usepackage{graphicx}
\usepackage{mathrsfs}
\usepackage{amssymb,amsfonts}
\usepackage[x11names]{xcolor}
\usepackage{enumerate}
\usepackage{xfrac}
\usepackage[normalem]{ulem}

\usepackage{geometry}
\hypersetup{pdftitle={Flavors of Astrophysical Neutrinos with Active--Sterile Mixing},
pdfsubject={Flavors of Astrophysical Neutrinos with Active--Sterile Mixing},
pdfauthor={Markus Ahlers, Mauricio Bustamante, and Niels Gustav Nortvig Willesen},
pdfstartview={FitH},
colorlinks=true,
bookmarksopen=false,
bookmarksnumbered=false,
bookmarksopenlevel=0,
linkcolor=Blue1!60!black,
citecolor=Green1!50!black,
urlcolor=Blue1!70!black}

\begin{document}

\title{Flavors of Astrophysical Neutrinos with Active-Sterile Mixing}
\author[a,1]{Markus Ahlers,\note{ORCID: \href{https://orcid.org/0000-0003-0709-5631}{0000-0003-0709-5631}}}
\emailAdd{markus.ahlers@nbi.ku.dk}
\author[a,b,2]{Mauricio Bustamante\note{ORCID: \href{https://orcid.org/0000-0001-6923-0865}{0000-0001-6923-0865}}
}
\emailAdd{mbustamante@nbi.ku.dk}
\author[a,3]{and\\Niels Gustav Nortvig Willesen\note{ORCID: \href{https://orcid.org/0000-0003-3532-9976}{0000-0003-3532-9976}}}
\emailAdd{wtb471@ku.dk}
\affiliation[a]{Niels Bohr International Academy, Niels Bohr Institute, University of Copenhagen, Blegdamsvej 17, DK-2100 Copenhagen, Denmark}
\affiliation[b]{DARK, Niels Bohr Institute, University of Copenhagen, Blegdamsvej 17, DK-2100 Copenhagen, Denmark}

\abstract{We revisit the flavor composition of high-energy astrophysical neutrinos observed at neutrino telescopes. Assuming unitary time evolution of the neutrino flavor states, the flavor composition observable at Earth is related to the initial composition at their sources via oscillation-averaged flavor transitions. In a previous study we derived general bounds on the flavor composition of TeV--PeV astrophysical neutrinos assuming three-flavor unitary mixing. We extend these bounds to the case of active-sterile neutrino mixing. Our bounds are analytical, derived based only on the unitarity of the mixing, and do not require sampling over the values of the unknown active-sterile mixing parameters. These bounds apply to any extended active-sterile neutrino mixing scenario where energy-dependent nonstandard flavor mixing dominates over the standard mixing observed in accelerator, reactor, and atmospheric neutrino oscillations.}


\arxivnumber{2009.01253}

\maketitle
\flushbottom

\section{Introduction}

The high-energy astrophysical neutrinos observed by IceCube~\cite{Aartsen:2013bka,Aartsen:2013jdh,Aartsen:2014gkd,Aartsen:2015rwa,Aartsen:2016xlq,IceCube:2018dnn,IceCube:2018cha} provide a unique probe of fundamental neutrino properties under extreme conditions. These neutrinos have energies of TeV--PeV and travel distances of up to a few Gpc, far exceeding those accessible to reactor, accelerator, or atmospheric neutrino experiments, that make them susceptible to tiny effects of nonstandard high-energy neutrino physics~\cite{Anchordoqui:2005is,Anchordoqui:2013dnh,Ahlers:2018mkf,Ackermann:2019cxh}. The observable features that could reveal the presence of nonstandard physics~\cite{Ackermann:2019cxh,Arguelles:2019rbn} include alterations of their energy spectrum, arrival direction distribution, arrival times, and flavor composition, {\it i.e.,} the proportion of neutrinos of each flavor in the neutrino flux. The latter is a particularly robust measure of nonstandard physics, since there are clear expectations for what the standard flavor composition should be, as predicted by Standard-Model interactions and oscillations between only the three active neutrino flavors, $\nu_e$, $\nu_\mu$, and $\nu_\tau$.

\begin{figure}[tbp]\centering
\includegraphics[width=0.6\linewidth,viewport=45 30 400 360,clip=true]{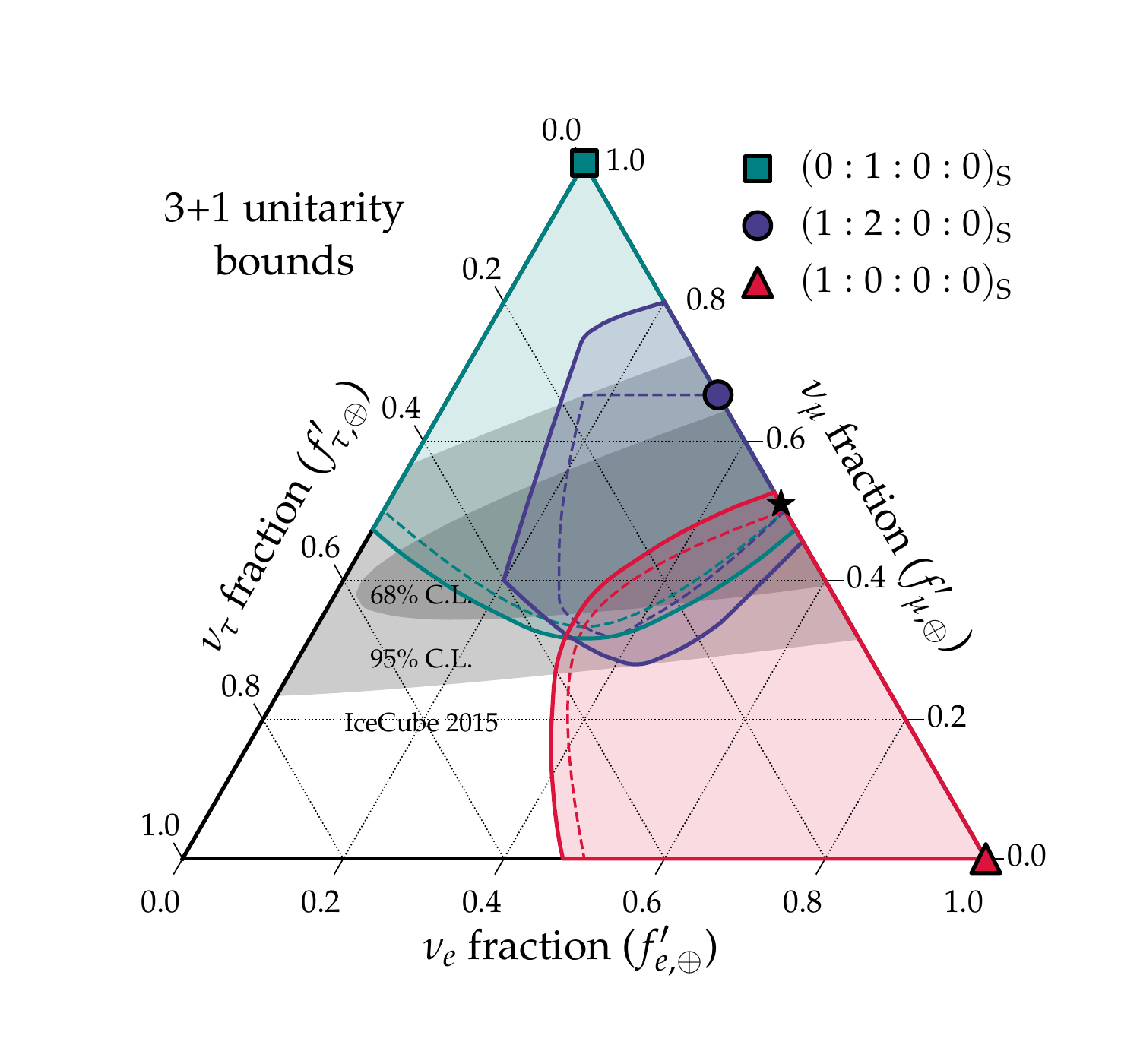}
\caption[]{Unitarity bounds of high-energy astrophysical neutrino flavors for three benchmark flavor compositions at the source indicated by filled symbols. The solid lines and shaded areas show the case of 3+1 active-sterile neutrino mixing, whereas the dashed lines show the case of mixing only between the three active flavors~\cite{Ahlers:2018yom}. We include the best-fit flavor composition measured by IceCube~\cite{Aartsen:2015knd} as a black star, and the $68\%$ and $95\%$ confidence levels (C.L.) as grey-shaded areas.}\label{fig1}
\end{figure}

High-energy astrophysical neutrinos are produced in interactions of high-energy cosmic rays, {\it i.e.}, protons and nuclei,  with gas and radiation in astrophysical sources.  (The identity of these sources remains so far unknown, save for two promising instances~\cite{IceCube:2018cha, Stein:2020xhk}.) The relative number of initial neutrino flavor states is determined by the physical conditions in the source and is dominated by electron and muon neutrinos $\nu_e$, $\bar{\nu}_e$, $\nu_\mu$, and $\bar{\nu}_\mu$. After emission from the source, neutrinos undergo flavor oscillations en route to Earth, which change the flavor composition with which they arrive at the detector~\cite{Beacom:2003nh, Kashti:2005qa, Xing:2006uk, Lipari:2007su, Pakvasa:2007dc, Esmaili:2009dz, Lai:2009ke, Choubey:2009jq}. Assuming standard three-flavor neutrino oscillations, the detected flavor composition can be corrected for these oscillation effects to infer the flavor composition at the source~\cite{Bustamante:2019sdb}.

However, nonstandard neutrino oscillations can alter the flavor composition at Earth drastically~\cite{Bustamante:2010bf, Xu:2014via, Fu:2014isa, Bustamante:2015waa, Gonzalez-Garcia:2016gpq, Nunokawa:2016pop, Rasmussen:2017ert,Ahlers:2018yom}. These effects can originate from a large class of models of new unitary neutrino physics, {\it e.g.}, from neutrino interactions with background matter~\cite{Bustamante:2018mzu}, dark matter~\cite{deSalas:2016svi,Capozzi:2018bps} or dark energy~\cite{Klop:2017dim,Capozzi:2018bps} or from Standard Model extensions that violate the weak equivalence principle, Lorentz invariance, or CPT symmetry~\cite{DeSabbata:1981ek, Gasperini:1989rt, Glashow:1997gx, Barenboim:2003jm, Bustamante:2010nq, Esmaili:2014ota, Arguelles:2015dca, Lai:2017bbl}. A key common property of these models is that, if the nonstandard effects dominate at high energies, the flavor transitions between the sources and Earth are entirely determined by a new unitary mixing matrix that connects neutrino flavor states and new, nonstandard propagation eigenstates that are motivated by these models~\cite{Akhmedov:2017mcc}. The values of the new mixing parameters, {\it i.e.}, the elements of the mixing matrix, are unknown or weakly constrained.  Naively, this complicates predicting the space of  nonstandard flavor compositions that we could expect at Earth. (A different class of models, which we do not study here, involves the nonunitary propagation of astrophysical neutrinos in scenarios with decoherence effects~\cite{Hooper:2004xr, Hooper:2005jp, Anchordoqui:2005gj, Bhattacharya:2010xj, Mehta:2011qb, Stuttard:2020qfv}.)

To overcome this issue, in Ref.~\cite{Ahlers:2018yom} we analytically derived the accessible space of flavor compositions that can be expected from this class of models, assuming oscillations between the three active flavors. Using the unitarity of the three-flavor lepton mixing matrix, we derived the boundary of the region that encloses all possible flavor compositions at the Earth for an arbitrary flavor composition at the source, {\it in spite of not knowing the values of the matrix elements}. These regions can be used to refine the search for nonstandard unitary physics in neutrino telescopes in an unbiased way.

Motivated by the possible existence of neutrino species beyond the three active flavors~\cite{Abazajian:2012ys, Kopp:2013vaa, Adhikari:2016bei, Boyarsky:2018tvu, Giunti:2019aiy, Diaz:2019fwt}, we revisit and expand this analysis by introducing mixing between the three active neutrinos and one sterile neutrino, {\it i.e.}, so-called 3+1 scenarios. Even if the active-sterile mixing is small, neutrino propagation over cosmological distances may affect the flavor composition at Earth appreciably.  In the context of high-energy astrophysical neutrinos, these extended flavor sectors have been previously investigated using particular values of an extended set of mixing parameters or by Monte-Carlo sampling them~\cite{Dutta:2001sf,Athar:2000yw,Awasthi:2007az,Donini:2008xn,Brdar:2016thq,Rasmussen:2017ert,Arguelles:2019tum}. Here, for the first time, we provide analytic flavor boundaries based on 3+1 unitarity constraints for arbitrary flavor compositions at the source. While there are numerous alternative proposals for the mass of the sterile neutrino, from eV to EeV, our flavor boundaries depend only indirectly on its mass and have wide applicability in testing active-sterile mixing.

Figure \ref{fig1} shows examples of our results for physically motivated choices of source flavor compositions. To visualize flavor boundaries in the ternary plot, we normalize each flavor contribution to the total number of active neutrinos after oscillation. The solid contours show our new 3+1 unitarity bounds; the dashed contours show the three-flavor neutrino bounds from our previous study~\cite{Ahlers:2018yom}.

The paper is organized as follows. In Sec.~\ref{sec1} we discuss the astrophysical processes of neutrino production and the corresponding flavor composition at the source. We discuss the resulting flavor composition at Earth after flavor oscillations with nonstandard 3+1 mixing. In Sec.~\ref{sec2} we derive general boundaries for the flavor composition at Earth based on the unitary of the 3+1 mixing matrix. We discuss our findings in Sec.~\ref{sec3} before we conclude in Sec.~\ref{sec4}. Throughout this paper we work in natural units with $\hbar=c=1$.

\section{Astrophysical Neutrino Flavors}\label{sec1}

We briefly review standard production mechanisms of high-energy astrophysical neutrinos. In astrophysical neutrino sources, cosmic-ray collisions with gas and radiation produce short-lived intermediate particles that, upon decaying, produce a flux of neutrinos $\nu_\alpha$ and antineutrinos $\overline\nu_\alpha$, where $\alpha = e, \mu, \tau$ refers to the active neutrino flavor eigenstate produced in weak interactions. In the 3+1 flavor scenario we introduce an additional sterile state denoted as $\nu_s$ that is not produced in weak interactions, but that mixes with the active neutrinos. The relative number of initial neutrino states $(N_e$\,:\,$N_\mu$\,:\,$N_\tau$\,:\,$N_s)_{\rm S}$ (summed over neutrinos and antineutrinos) is determined by the physical conditions in the source. For astrophysical sources we expect an initial flavor composition dominated by $\nu_e$ and $\nu_\mu$ with only little contribution from $\nu_\tau$~\cite{Athar:2005wg}. Sterile neutrino production at the source can only be a result of physics beyond the Standard Model~\cite{Brdar:2016thq}, {\it e.g.}, from the decay and/or annihilation of dark matter. 

In the simplest case, pions (or kaons) produced in cosmic-ray interactions decay via $\pi^+\to\mu^++\nu_\mu$ followed by $\mu^+\to e^++\nu_e+\overline\nu_\mu$ (and the charge-conjugated processes). This pion decay chain results in a source composition of $($1\,:\,2\,:\,0\,:\,0$)_{\rm S}$. However, in the presence of strong magnetic fields it is possible that muons lose energy before they decay and do not contribute to the high-energy neutrino emission~\cite{Kashti:2005qa}. In this muon-damped scenario the composition is expected to be closer to $($0\,:\,1\,:\,0\,:\,0$)_{\rm S}$. On the other hand, neutrino production by beta-decay of free neutrons or short-lived isotopes produced in spallation or photo-disintegration of cosmic rays leads to $($1\,:\,0\,:\,0\,:\,0$)_{\rm S}$. Below, we use these three physically motivated cases as benchmarks.

After production, astrophysical neutrinos travel over cosmic distances before their arrival at Earth. The flavor composition at Earth is significantly altered by neutrino oscillations, which are due to each neutrino flavor state being a superposition of propagation states $\nu_\mathfrak{a}$ ($\mathfrak{a} = 1,2,3,4$),
\begin{equation}\label{eq:Uneutrino}
  |\nu_\alpha\rangle =
  \sum_\mathfrak{a} U_{\alpha \mathfrak{a}}^* |\nu_\mathfrak{a}\rangle\,,
\end{equation}
where $\alpha = e,\mu,\tau,s$, and $U_{\alpha \mathfrak{a}}$ is an element of the mixing matrix $\bf U$ that connects the flavor and propagation states. The propagation states are defined as eigenvectors of the Hamiltonian, including kinetic terms and effective potentials~\cite{Akhmedov:2017mcc}.  In the standard three-flavor scenario, $\bf U$ is the {\it Pontecorvo-Maki-Nakagawa-Sakata} $3\times3$ unitary matrix~\cite{Pontecorvo:1957qd,Maki:1962mu,Pontecorvo:1967fh} with 6 degrees of freedom --- 3 mixing angles and 3 physical phases.  In the 3+1 scenario, in general, $\bf U$ is a $4\times4$ unitary mixing with 12 degrees of freedom --- 6 mixing angles and 6 physical phases. However, three (Majorana) phases do not affect neutrino oscillations. Unitarity ensures that the total number of neutrinos of all flavors is conserved. Neutrino flavor oscillations of pure or mixed states can be described in terms of the evolution of the density matrix $\rho$, following the Liouville equation $\dot\rho= -{\rm i}[H,\rho]$ with Hamiltonian $H$.

On their way to Earth, high-energy astrophysical neutrinos propagate in vacuum. Usually, the propagation eigenstates are the neutrino mass eigenstates $\nu_i$ ($i=1,2,3,4$). For illustration, in 3+1 models where the active-sterile mixing parameters are small, $\nu_1$, $\nu_2$, and $\nu_3$ are made up mostly of the active flavors, $\nu_e$, $\nu_\mu$, and $\nu_\tau$, with a small contribution of the sterile flavor $\nu_s$, while $\nu_4$ is mostly made up $\nu_s$, with a small contribution of the active flavors.  However, our treatment below is not limited to the case of small active-sterile mixing; it holds regardless of the size of the mixing parameters.

If neutrinos are relativistic, like in our case, standard oscillations in vacuum can be introduced via the Hamiltonian
\begin{equation}\label{eq:H0}
H_0 \simeq \sum_{i=1}^4\bigg(p + \frac{m_i^2}{2p}\bigg)\big(|\nu_i\rangle\langle\nu_i|+|\overline{\nu}_i\rangle\langle\overline{\nu}_i|\big)\,,
\end{equation}
where $p\simeq E$ is the neutrino momentum and the sum runs over projectors onto neutrino and antineutrino mass eigenstates. The solution of the Liouville equation then yields the probability of transition between neutrino flavors due to their mixing, coming from $\bf U$, and their mass splittings, $\Delta m_{ij}^2 \equiv m_i^2 - m_j^2$, where $i,j=1,2,3,4$. Formally, the probability is oscillatory, and the oscillation phases are given by $\Delta m_{ij}^2\ell/4E$ where $\ell$ is the distance to the neutrino source. 

Global analyses of oscillation data from reactor, solar, and atmospheric neutrino experiments~\cite{Capozzi:2018ubv,Esteban:2020cvm}, with energies in the MeV--GeV scale, confirm the validity of the three-flavor oscillation phenomenology. However, there is motivation from theory and experiment to consider the existence of an additional, sterile neutrino.  From theory, sterile neutrinos appear naturally in the process of giving masses to the active neutrinos; see, {\it e.g.}, Refs.~\cite{Bilenky:1998dt, Mohapatra:2006gs}. These sterile neutrinos are typically very heavy; for instance, in type-I seesaw models of mass generation, they have masses of $\sim 10^{25}$~eV.  From experiment, eV-scale sterile neutrinos~\cite{Abazajian:2012ys, Kopp:2013vaa, Giunti:2019aiy, Diaz:2019fwt} are motivated by hints from the short-baseline oscillation experiments LSND~\cite{Aguilar:2001ty} and MiniBooNE~\cite{AguilarArevalo:2007it,Aguilar-Arevalo:2013pmq}, from the Gallium neutrino anomaly~\cite{Giunti:2006bj,Giunti:2010zu,Giunti:2012tn}, and from anomalies in reactor neutrino experiments~\cite{Mention:2011rk}, while keV-scale sterile neutrinos are motivated as dark-matter candidates~\cite{Adhikari:2016bei,Boyarsky:2018tvu} by astrophysical X-ray observations~\cite{Bulbul:2014sua}.  

Concurrently, and in tension with these hints, there are strong experimental constraints, derived from MeV--GeV oscillation experiments and GeV--TeV atmospheric neutrino observations, that limit active-sterile mixing to be small~\cite{Giunti:2019aiy,Diaz:2019fwt,Abe:2014gda,Aartsen:2017bap,Aartsen:2020iky,Aartsen:2020fwb}. If these constraints were to hold also for active-sterile mixing at the TeV--PeV scale, only small deviations would be possible in the flavor composition at Earth of high-energy astrophysical neutrinos~\cite{Brdar:2016thq,Arguelles:2019tum} (unless $\nu_s$ were produced at the sources). In our treatment, we make no such assumption: we allow active-sterile mixing in the TeV--PeV scale to be disconnected from active-sterile mixing in the MeV--TeV scale. This approach can be motivated, {\it e.g.}, by Lorentz invariance violating (LIV) extensions\footnote{For simplicity, we will here only consider isotropic CPT-even terms that affect neutrinos and anti-neutrinos equally. Note that the unitarity boundaries derived in this paper do not depend on a specific form of LIV. However, our boundaries are particularly relevant for LIV Hamiltonians with a strong energy dependence that can facilitate nonstandard mixing towards PeV energies. For a general classification of LIV Hamiltonians we refer to Ref.~\cite{Kostelecky:2011gq}.} of the conventional neutrino Hamiltonian (\ref{eq:H0}) in the form~\cite{Kostelecky:2011gq}
\begin{equation}\label{eq:Heff}
\Delta H_{\rm LIV} =  \bigg(\frac{p}{\Lambda}\bigg)^n \sum_{\mathfrak{a}} \epsilon_\mathfrak{a}\big(|\nu_{\mathfrak{a}}\rangle\langle\nu_{\mathfrak{a}}| + |\overline{\nu}_{\mathfrak{a}}\rangle\langle\overline{\nu}_{\mathfrak{a}}|\big) \,,
\end{equation}
with {\it odd} integer $n\geq1$. The eigenstates $|\nu_\mathfrak{a}\rangle$ and $|\overline\nu_\mathfrak{a}\rangle$ of Eq.~(\ref{eq:Heff}) are unrelated to the mass eigenstates $|\nu_i\rangle$ and $|\overline\nu_i\rangle$ that appear in Eq.~(\ref{eq:H0}). The eigenvalues of $\Delta H_{\rm LIV}$ are required to be non-degenerate, {\it i.e.}, $\Delta\epsilon_{\mathfrak{a}\mathfrak{b}} \equiv \epsilon_\mathfrak{a} - \epsilon_\mathfrak{b} \neq 0$ for $\mathfrak{a}\neq\mathfrak{b}$, in order to induce neutrino oscillations.

Due to its strong energy dependence, the Hamiltonian term $\Delta H_{\rm LIV}$ can dominate neutrino flavor oscillations at high energies, while low-energy neutrino phenomena remain essentially unaffected. In this case, the eigenstates $\nu_\mathfrak{a}$ and $\overline\nu_\mathfrak{a}$ of Eq.~(\ref{eq:Heff}) form the relevant basis to describe neutrino mixing at high energies, with oscillation phases given by $\Delta\epsilon_{\mathfrak{a}\mathfrak{b}}(E/\Lambda)^n\ell/2$. In the following, and motivated by earlier work~\cite{Arguelles:2015dca,Ahlers:2018yom,Arguelles:2019tum}, we will assume that this situation applies to IceCube's neutrino observation in the 10~TeV--10~PeV energy range. Assuming that the transition between different oscillation regimes occurs at an intermediate neutrino energy $E^*\simeq 1$~TeV we can relate the size of the 3+1 mass splittings to the LIV energy levels as ${\rm max}(\Delta m_{ij}^2)\simeq m^2_4\simeq{\rm min}(\Delta\epsilon^*_{\mathfrak{a}\mathfrak{b}})\times1~{\rm TeV}$, where $\Delta\epsilon^*_{\mathfrak{a}\mathfrak{b}}$ is the difference between eigenvalues of $\Delta H_{\rm LIV}$, Eq.~(\ref{eq:Heff}), at $E^*\simeq1$~{\rm TeV}. This condition can be met by a wide range of sterile neutrino masses and LIV energy splittings.

Due to the large distance of astrophysical neutrino sources it is reasonable to assume that high-energy oscillation phases are large, {\it i.e.}, that oscillations are rapid. In this case, considering the wide energy distribution with which neutrinos are emitted and the limited energy resolution of neutrino detectors, flavor transitions from $\nu_\alpha$ to $\nu_\beta$ (or from $\overline\nu_\alpha$ to $\overline\nu_\beta$) can only be described by their oscillation-averaged transition probability, given by~\cite{Learned:1994wg, Athar:2000yw}
\begin{equation}\label{eq:Paverage}
{P}_{\alpha\beta} = \sum_{\mathfrak{a}=1}^4|U_{\alpha \mathfrak{a}}|^2\, |U_{\beta \mathfrak{a}}|^2\,.
\end{equation}
Here, the unitary matrix $U$ describes the mixing of 3+1 flavor states $\nu_\alpha$ with the high-energy propagation states $\nu_\mathfrak{a}$.

The transition probability, Eq.~(\ref{eq:Paverage}), allows us to study the generic situation where the mixing parameters are different in different energy regimes, or where different sterile flavors mix with active flavors at different energies. As a result, we circumvent the constraints on the active-sterile mixing coming from MeV--TeV experiments, and base our constraints on the flavor composition at Earth below solely on the unitarity of the mixing matrix. Later, we point out how to use our flavor-composition constraints to indirectly probe the active-sterile mixing parameters.

\section{Flavor Boundaries}\label{sec2}

Our goal is to derive the boundary that encloses the accessible space of flavor compositions at Earth, for a given flavor composition at the source, based solely on the unitarity of the $4\times4$ mixing matrix $\bf U$. The derivation of flavor boundaries in the presence of sterile neutrinos follows closely that of the three-flavor case described in Ref.~\cite{Ahlers:2018yom}. We focus only on the 3+1 scenario, which contains a single sterile flavor, because it is representative of the class of 3+$n$ scenarios, with $n \geq 1$ sterile neutrinos.  Our formalism below can be extended to scenarios with $n > 1$.   

In the 3+1 scenario, the oscillation-averaged flavor transition matrix ${\bf P}$ defined by Eq.~(\ref{eq:Paverage}) can be parametrized by its six off-diagonal entries $P_{\alpha\beta}$, with $\alpha \neq \beta$. The unitarity of the mixing matrix imposes a bound on the linear combinations of these transition elements,
\begin{equation}\label{eq:bound}
u{P}_{es} + v {P}_{\mu s} + w{P}_{\tau s} + x{P}_{\mu\tau} + y {P}_{e\tau} + z{P}_{e\mu} 
\leq B(u,v,w,x,y,z)\,,
\end{equation}
where $u$, $v$, $w$, $x$, $y$, and $z$ are arbitrary parameters and $B$ is the boundary function. We discuss its form in Appendix~\ref{appI}. This function can be written as
\begin{equation}\label{eq:B}
B(u,v,w,x,y,z) = \max \bigg(\bigcup_{i=1}^{40}\mathcal{S}_i\bigg)\,,
\end{equation}
where the individual subset $\mathcal{S}_i$ corresponds to a class of candidate maxima of the left-hand side of Eq.~(\ref{eq:bound}) that are related to one another by flavor transformations. In total, we consider 40 classes of candidate maxima that are listed in Table~\ref{tab1} of Appendix~\ref{appI}.

As in the three-flavor case~\cite{Ahlers:2018yom}, it is possible to use the family of unitarity bounds in Eq.~(\ref{eq:G}) of Appendix~\ref{appI} to derive boundaries that enclose the accessible region of flavor compositions at Earth. We first define the flavor ratio of $\nu_\alpha$ as $f_{\alpha} \equiv N_\alpha/\sum_\beta N_\beta$. For a fixed source flavor ratio $f_{\alpha,{\rm S}}$,   the flavor ratio at Earth $f_{\alpha,\oplus}$ is
\begin{equation}
f_{\alpha,\oplus} = \sum_\beta{P}_{\alpha\beta}f_{\beta,{\rm S}}\,.
\end{equation}
In the trivial case where there is no mixing, ${\bf U}={\bf I}$, the oscillation-averaged transition probability is also trivial, {\it i.e.}, $P_{\alpha\beta} = \delta_{\alpha\beta}$, and so $f_{\alpha,\oplus}=f_{\alpha,{\rm S}}$. Therefore, the original flavor composition at the source is always part of the accessible space of flavor composition at Earth. Since there is a continuous parametrization of the transition matrix ${\bf P}$ in terms of mixing angles and phases, the accessible space of flavor composition at Earth must be connected (although not necessarily simply connected). This means that, within that space, it is possible to transform continuously between $f_{\alpha, \oplus} = f_{\alpha, {\rm S}}$ and any other flavor composition. Therefore, to find the boundary that encloses the accessible flavor space at Earth, we look for the boundary of the flavor shift defined as 
\begin{equation}
\Delta f_{\alpha} \equiv f_{\alpha,\oplus} - f_{\alpha,{\rm S}}\,.
\end{equation}
Due to 3+1 unitarity, we have $\sum_\alpha \Delta f_{\alpha} =0$ and can therefore parametrize the total flavor shift by only three parameters, which we choose to be $\Delta f_e$, $\Delta f_\mu$, and $\Delta f_s$. 

We can now look at hyper-surfaces in the three-dimensional flavor space of $\Delta f_e$, $\Delta f_\mu$, and $\Delta f_s$, defined via $\widehat{\bf n}\cdot(\Delta f_e,\Delta f_\mu,\Delta f_s) = \text{const.}$, where $\widehat{\bf n}$ is a three-dimensional unit vector in an arbitrary direction, which we vary later in order to scan the flavor space in all directions. The projection of the flavor shift $(\Delta f_e,\Delta f_\mu,\Delta f_s)$ onto $\widehat{\bf n}$ can be written as on the left-hand side of Eq.~(\ref{eq:bound}) with the coefficients
\begin{align}
u &= (f_{e,{\rm S}} - f_{s,{\rm S}}) (\widehat{n}_s - \widehat{n}_e)\,,\\
v &= (f_{\mu,{\rm S}} - f_{s,{\rm S}}) (\widehat{n}_s - \widehat{n}_\mu)\,,\\
w &= (1 - f_{e,{\rm S}} - f_{\mu,{\rm S}} - 2f_{s,{\rm S}}) \widehat{n}_s\,,\\
x &= (1 - f_{e,{\rm S}} - 2f_{\mu,{\rm S}} - f_{s,{\rm S}}) \widehat{n}_\mu\,,\\
y &= (1 - 2f_{e,{\rm S}} - f_{\mu,{\rm S}} - f_{s,{\rm S}}) \widehat{n}_e\,,\\
z &= (f_{e,{\rm S}} - f_{\mu,{\rm S}}) (\widehat{n}_\mu - \widehat{n}_e)\,.
\end{align}
In other words, given an arbitrary direction $\widehat{\bf n}$ in flavor space, the boundary in Eq.~(\ref{eq:bound}) translates into a hyper-surface boundary in flavor space in that direction. 

Because neutrino telescopes detect only active flavors, we project the three-dimensional hyper-surface boundaries in the flavor space of $\Delta f_e$, $\Delta f_\mu$, and $\Delta f_s$ onto the corresponding boundaries in the two-dimensional subspace of active flavors. To connect to the observations of neutrino telescopes, we define the flavor fraction of active neutrinos, $f'_\alpha$, which is related to the flavor fraction of all neutrinos, $f_\alpha$, as
\begin{equation}
f'_\alpha \equiv \frac{f_\alpha}{1-f_s}\,.
\end{equation}
With this definition we have $f'_e +f'_\mu +f'_\tau = 1$ and, for a given arbitrary source flavor composition, we are able to derive boundaries in the subspace of active flavor fractions $f_e^\prime$ and $f_\mu^\prime$, and show them in a ternary plot. The procedure is outlined in Appendix~\ref{appII}. 

\begin{figure*}[tbp]\centering
\includegraphics[width=0.47\linewidth,viewport=45 30 400 365,clip=true]{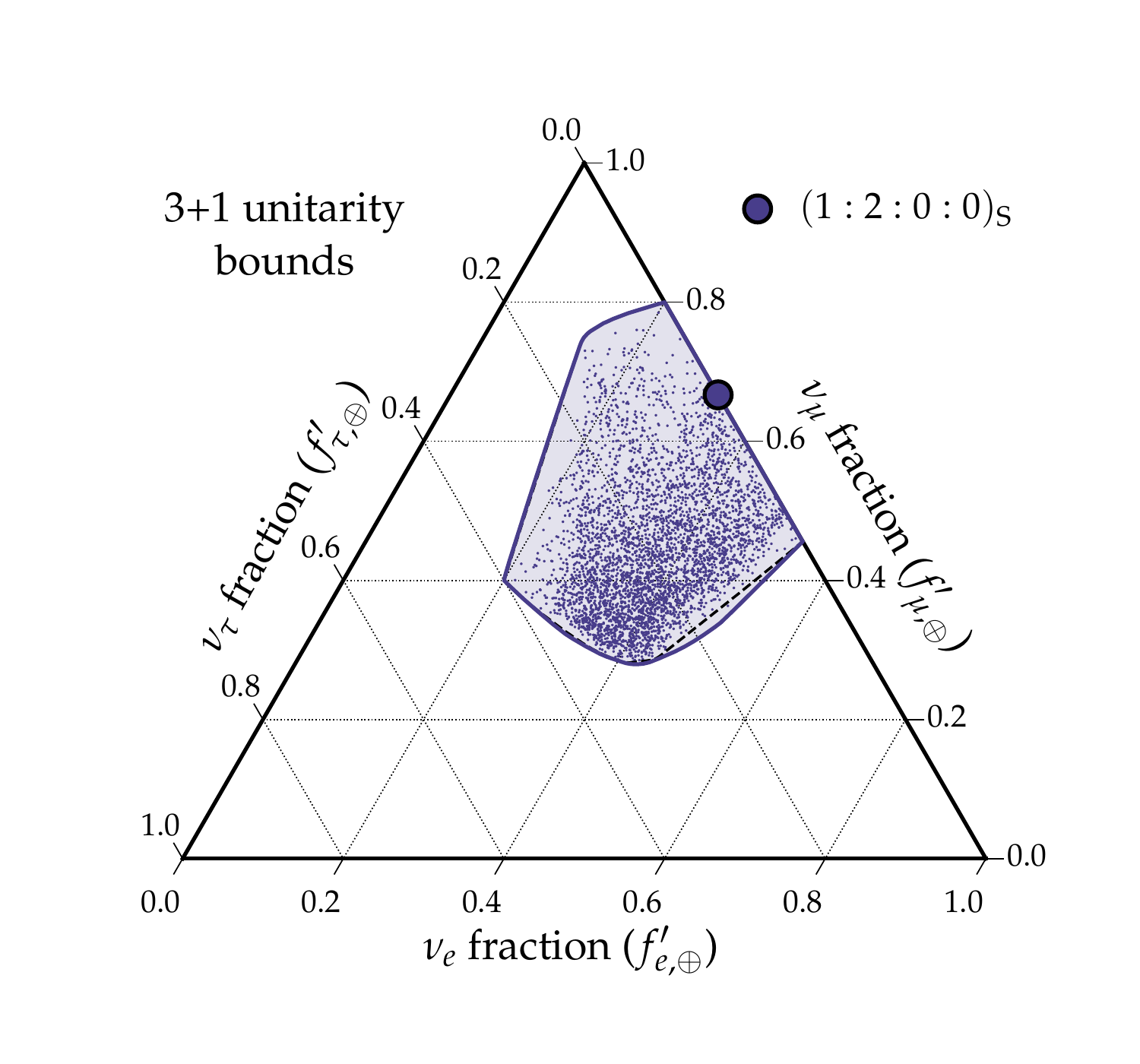}\hspace{0.3cm}\includegraphics[width=0.47\linewidth,viewport=45 30 400 365,clip=true]{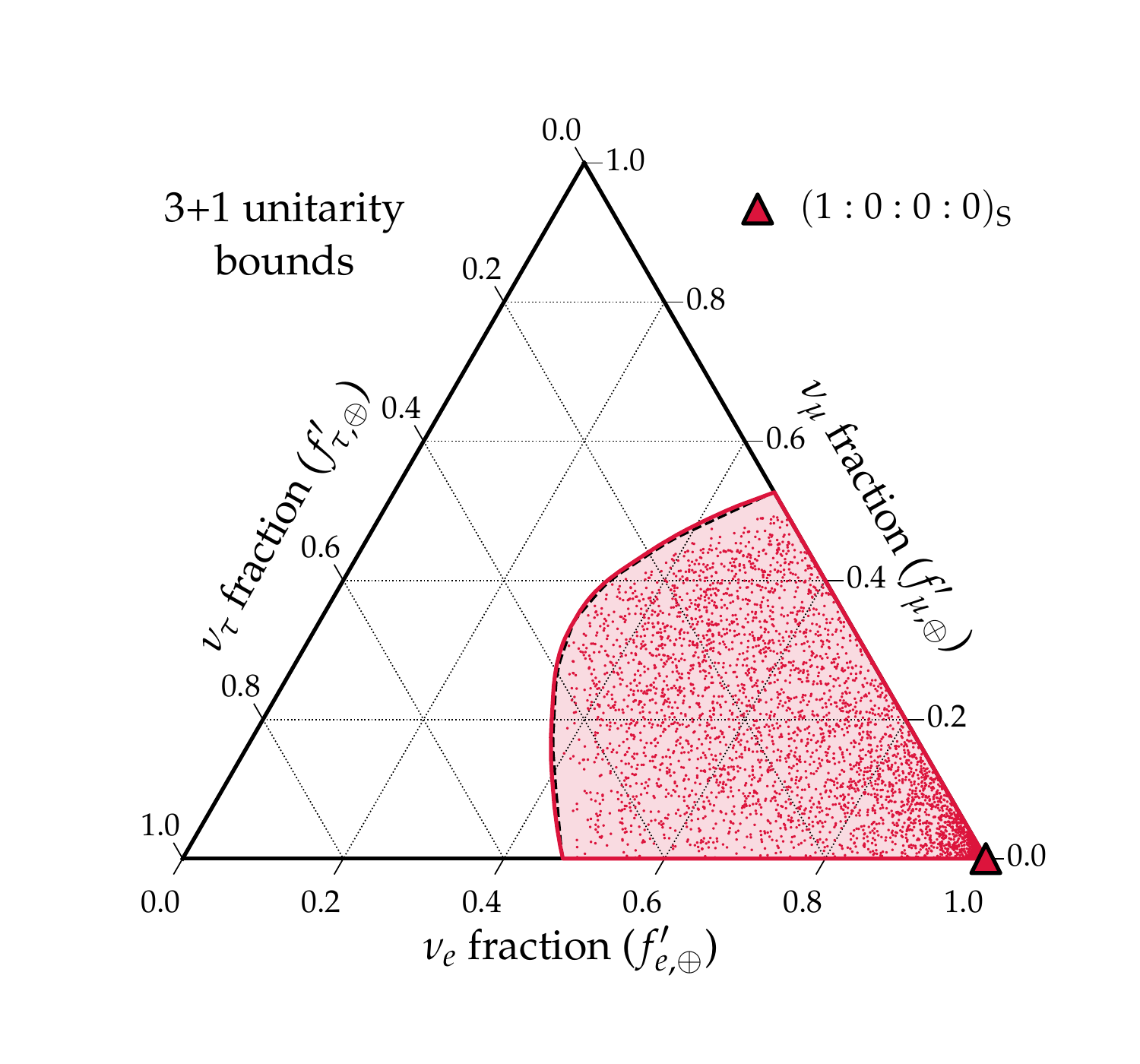}
\caption[]{Comparison of unitarity bounds vs.~the active flavor fractions computed for 4,000~random realizations of the unitary mixing matrix. {\bf Left:} Unitarity bound (solid line) from Fig.~\ref{fig1} and random realizations for a source flavor composition of $($1$\,:\,$2$\,:\,$0$:\,$0$)_{\rm S}$. We also show the reduced boundary (dashed line) derived from the subset of local extrema in Eq.~(\ref{eq:B}) that obey the quadrilateral inequality; see the main text and Appendix~\ref{appI} for details. {\bf Right:} Same as in the left panel, but now showing the case $($1$\,:\,$0$\,:\,$0$:\,$0$)_{\rm S}$. This case is related to $($0$\,:\,$1$\,:\,$0$:\,$0$)_{\rm S}$ via index permutations, as described in Appendix~\ref{appI}.}\label{fig2}
\end{figure*}

Figure~\ref{fig1} shows the resulting boundaries of the accessible active flavor fractions at Earth for our three benchmark cases of flavor composition at the source. The gray-shaded areas in Fig.~\ref{fig1} indicate the 68\% and 95\% confidence levels (C.L.s) from a flavor-composition analysis carried out by IceCube~\cite{Aartsen:2015knd}. Due to the difficulty in distinguishing between events induced by $\nu_e$ and $\nu_\tau$ in the IceCube data~\cite{Abbasi:2012cu,Aartsen:2015dlt}, the likelihood contour is presently rather flat along the $f_\mu$ direction, leading to almost horizontal confidence levels in the ternary plot of Fig.~\ref{fig1}~\cite{Mena:2014sja, Aartsen:2015ivb, Palomares-Ruiz:2015mka, Vincent:2016nut}. This degeneracy could be lifted with future data\footnote{Indeed, two recent IceCube analyses identified the first $\bar\nu_e$ candidate event via the Glashow resonance~\cite{IceCube:2021rpz} and $\nu_\tau+\bar\nu_\tau$ candidates by ``double cascades''~\cite{Abbasi:2020zmr,Abbasi:2020jmh}. The impact on the flavor likelihood will be apparent in future updates of IceCube's global analysis~\cite{Aartsen:2015knd} that we presently show in the ternary plots.} by the observation of characteristic $\bar{\nu}_e$~\cite{Glashow:1960zz, Anchordoqui:2004eb, Bhattacharya:2011qu, Barger:2014iua, Palladino:2015vna} and $\nu_\tau$ events~\cite{Learned:1994wg, Beacom:2003nh, Palladino:2018qgi, Aartsen:2020fgd}; see, {\it e.g.}, the flavor projections in Ref.~\cite{Song:2020nfh}.

Under the assumption of standard three-flavor oscillations, the measured flavor composition disfavors the source composition $($1\,:\,0\,:\,0$)_{\rm S}$~\cite{Bustamante:2019sdb}. However, the unitarity bounds in Fig.~\ref{fig1} indicate that there are nonstandard unitary oscillation scenarios in the three-flavor and 3+1 scenarios that can be consistent with the IceCube measurements within the $68\%$ C.L.

By construction, the boundary in Eq.~(\ref{eq:bound}) encloses a convex space and, therefore, its projection onto the subspace of active flavor fractions is a convex boundary, {\it i.e.}, one in which every line segment between any two points is contained in the subspace. It is a nontrivial question if every flavor combination within the boundary can be actually realized by at least one unitary mixing matrix. For the three-flavor mixing discussed in Ref.~\cite{Ahlers:2018yom}, we showed that our convex unitarity boundary in Eq.~(\ref{eq:bound}) accurately represents the accessible flavor space for the three benchmark production scenarios shown in Fig.~\ref{fig1}. For comparison, we show also these three-flavor boundaries in Fig.~\ref{fig1}. However, even in the simpler three-flavor case there are other flavor compositions at the source, different from the benchmark cases, for which the space of flavor composition at the Earth is not convex and, therefore, the boundary is not maximally constraining. The same is also true in the 3+1 scenario. 

In the case of 3+1 flavor mixing, an additional complication arises from the existence of local extrema that lie on the boundary implied by the quadrilateral inequalities that come from the off-diagonal elements of the unitarity condition ${\bf U}^\dagger{\bf U} = {\boldsymbol 1}$, Eq.~(\ref{eq:bound2}) in Appendix~\ref{appI}. We did not find a simple algorithmic way to derive these additional extrema within the formalism that we use to include the diagonal unitarity conditions, via Lagrange multipliers, as detailed in Appendix~\ref{appI}. However, including these additional extrema can only {\it shrink} the boundary in Eq.~(\ref{eq:bound}). Therefore, the boundary that we have derived without these additional extrema still fully encloses the accessible flavor space, even if it may not be the {\it minimal} convex boundary.

To gauge the importance of these missing extrema, we construct a reduced version of the boundary based on Eq.~(\ref{eq:bound}), but now considering only those extrema $\mathcal{S}_i$ that obey the quadrilateral inequality, Eq.~(\ref{eq:bound2}). The plots in Fig.~\ref{fig2} compare our full flavor boundary (solid lines) to the reduced boundary (dashed lines) for the source compositions $($1$\,:\,$2$\,:\,$0$:\,$0$)_{\rm S}$ (left plot) and $($1$\,:\,$0$\,:\,$0$:\,$0$)_{\rm S}$ (right plot). The {\it minimal} convex boundary based on the full set of extrema necessarily has to lie {\it between} these two boundaries. As can be seen, the effect of local extrema can only be marginal for these cases and our analytical boundary is already close to minimal.

Figure~\ref{fig2} also shows the distribution of observed flavor ratios $f'_\alpha$ from 4,000 random realizations of unitary mixing parameters for the individual source compositions. As expected, our analytical boundary derived using Eq.~(\ref{eq:bound}) fully encloses the scattered data while the reduced boundaries omit a few data points. This indicates that the {\it minimal} convex 3+1 flavor boundaries can depend on local extrema that saturate the quadrilateral inequality. Nevertheless, our method allows to set nontrivial boundaries on the active flavor fractions that closely trace the distribution of the data on the ternary plot.

\section{Discussion}\label{sec3}

Figure~\ref{fig1} summarizes the unitarity bounds for the three-flavor (dashed) and 3+1 scenarios (solid),  assuming three benchmark source production mechanisms, $($1\,:\,2\,:\,0\,:\,0$)_{\rm S}$, $($0\,:\,1\,:\,0\,:\,0$)_{\rm S}$, and $($1\,:\,0\,:\,0\,:\,0$)_{\rm S}$. The latter two are related via the exchange $f_e\leftrightarrow f_\mu$ and this symmetry is clearly visible in the corresponding boundaries. These results show that, if no $\nu_s$ are produced at the astrophysical sources, as predicted by standard neutrino production processes, the boundary of allowed flavor composition at Earth in the 3+1 scenariois extended enlarged compared to the three-flavor scenario. This is expected, as the three-flavor scenario is a special case of the 3+1 scenario with no mixing between active and sterile neutrinos.

The grey-shaded areas in Fig.~\ref{fig1} indicate the preferred regions of flavor composition measured by IceCube~\cite{Aartsen:2015knd} at the $68\%$ and $95\%$ confidence levels (C.L.s). As already discussed in Ref.~\cite{Ahlers:2018yom}, the three-flavor unitarity boundaries of the three benchmark production mechanisms have significant overlap with the preferred IceCube regions. In particular, in the three-flavor scenario, while $($1\,:\,0\,:\,0\,$)_{\rm S}$ is disfavored by IceCube under standard neutrino mixing, the unitarity boundary is consistent with IceCube at the 68\% C.L. In the 3+1 scenario, because the unitarity boundary is larger, the agreement with IceCube increases and the allowed flavor region also includes now the best-fit flavor composition from Ref.~\cite{Aartsen:2015knd}, indicated by a black star.

\begin{figure*}[tbp]\centering
\includegraphics[width=0.47\linewidth,viewport=45 30 400 365,clip=true]{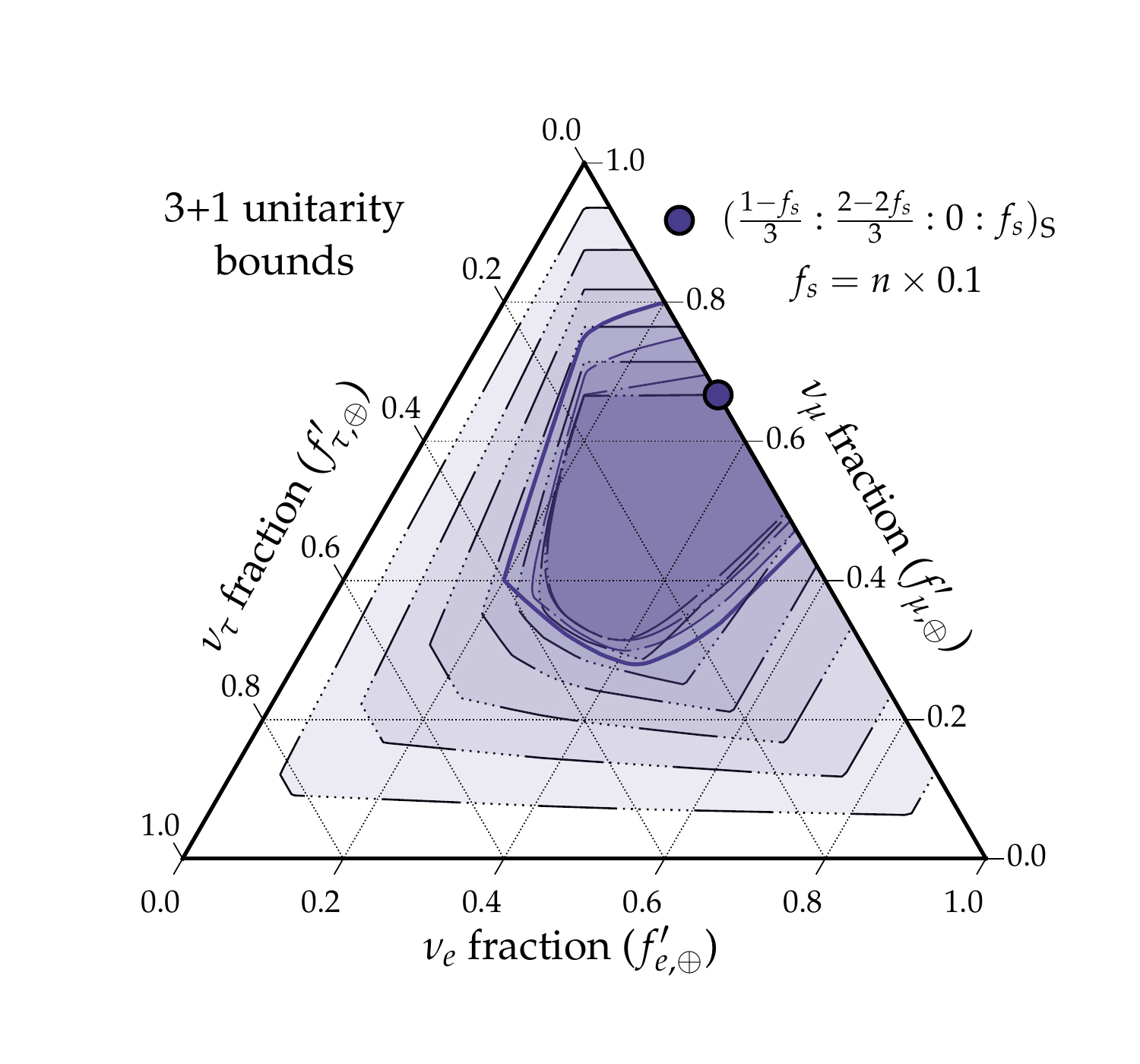}\hspace{0.3cm}\includegraphics[width=0.47\linewidth,viewport=45 30 400 365,clip=true]{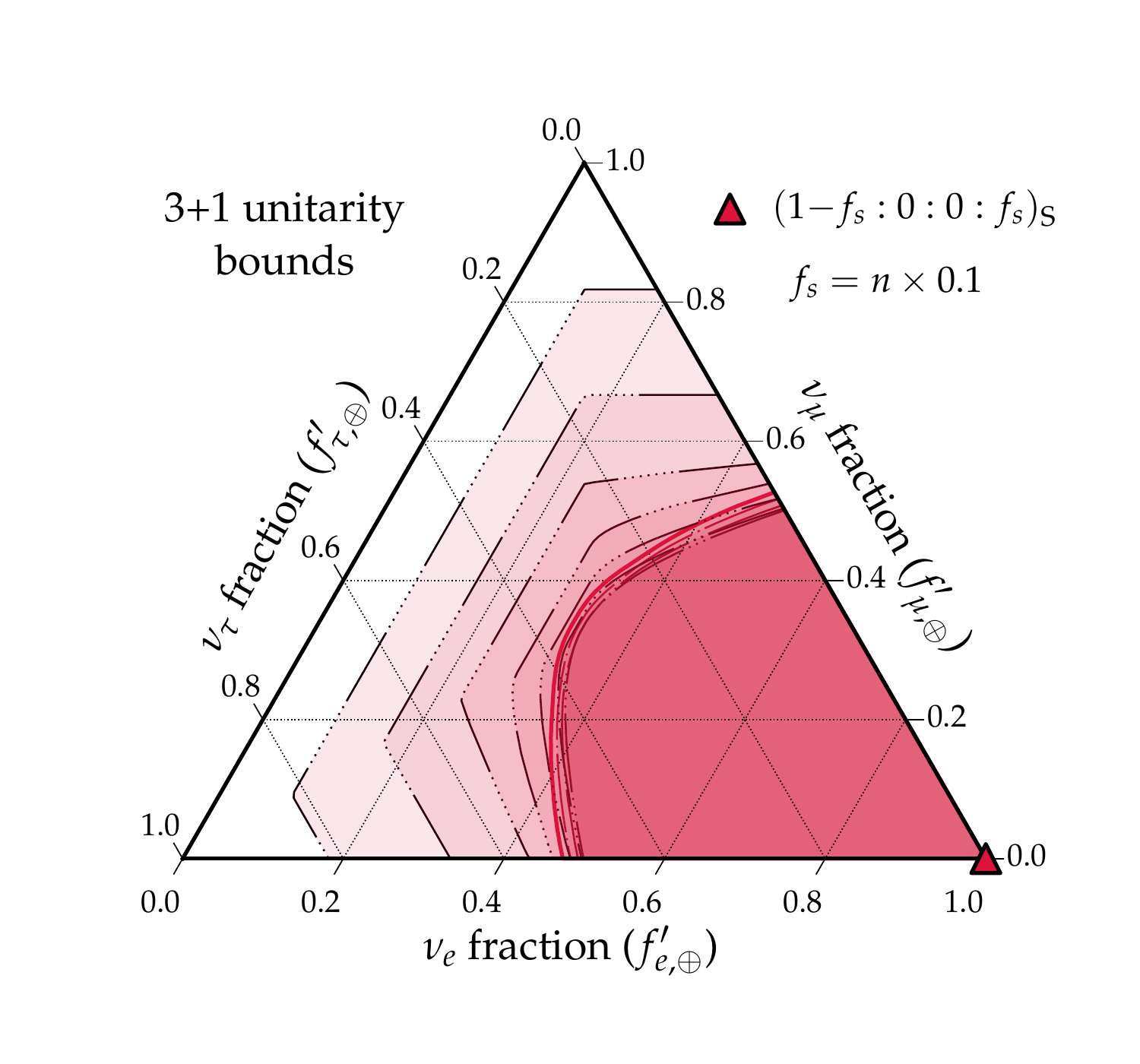}
\caption[]{Flavor boundaries for a source composition $((1-f_s)/3:(2-2f_s)/3:0:f_s)_{\rm S}$ (left) and $(1-f_s:0:0:f_s)_{\rm S}$ (right). The sterile fractions at the source is fixed as $f_{s,{\rm S}}=n\times0.1$ with integer $n$ indicated by the number of consecutive ``dots'' in the boundary lines and increasingly darker line colors. The case $f_{s,{\rm S}}=0$ corresponds to the boundary shown in Fig.~\ref{fig1}. With increasing $f_{s,{\rm S}}$, the boundaries first shrinks and then grows.}\label{fig3}
\end{figure*}

It is an interesting observation that in the 3+1 scenario, with 6 mixing angles and 3 Dirac phases, the unitarity boundaries are only marginally extended compared to the three-flavor scenario, with only four degrees of freedom. In particular, the union of the 3+1 flavor boundaries for source compositions $($0\,:\,1\,:\,0\,:\,0$)_{\rm S}$ and $($1\,:\,0\,:\,0\,:\,0$)_{\rm S}$ represent the boundary of the general case $(f_e:1-f_e:0:0)_{\rm S}$, {\it i.e.}, all possible standard astrophysical sources, where there is no significant production of $\nu_\tau$ and $\nu_s$. For these sources, even in the 3+1 scenario there are still flavor compositions at the 68\% C.L.~that are not covered by the extended unitarity boundary. If future IceCube data with higher precision prefers a tau neutrino fraction of $f'_{\tau,\oplus}\gtrsim0.53$, this will not be explainable by the 3+1 model and standard neutrino production mechanisms.

The situation changes if we consider the production of sterile neutrinos at the source. Figure~\ref{fig3} shows the effect on the flavor boundary of varying the sterile neutrino fraction at the source, $f_{s,{\rm S}}$, which explores nonstandard neutrino production mechanisms. There, we show the behavior of the boundary for a source flavor composition $((1-f_s)/3:(2-2f_s)/3:0:f_s)_{\rm S}$ (left plot) and $(1-f_s:0:0:f_s)_{\rm S}$ (right plot), where we increase $f_{s,{\rm S}}$ in steps of 10\%. As we increase $f_{s,{\rm S}}$, the contours first shrink until we reach $f_{s,{\rm S}} \simeq 0.3$, before growing for larger values of $f_{s,{\rm S}}$ until the boundary encloses the full flavor space at $f_{s,{\rm S}}=1$. This indicates that the accessible region of active flavor fractions at Earth is practically unconstrained if there is significant sterile neutrino production at the source, in agreement with the results from Refs.~\cite{Brdar:2016thq, Arguelles:2019tum}. Note that all individual 3+1 boundaries in Fig.~\ref{fig3} enclose the three-flavor boundary shown in Fig.~\ref{fig1}, corresponding to the trivial case of no mixing between active and sterile neutrinos.

\begin{figure*}[tbp]\centering
\includegraphics[width=0.47\linewidth,viewport=45 30 400 365,clip=true]{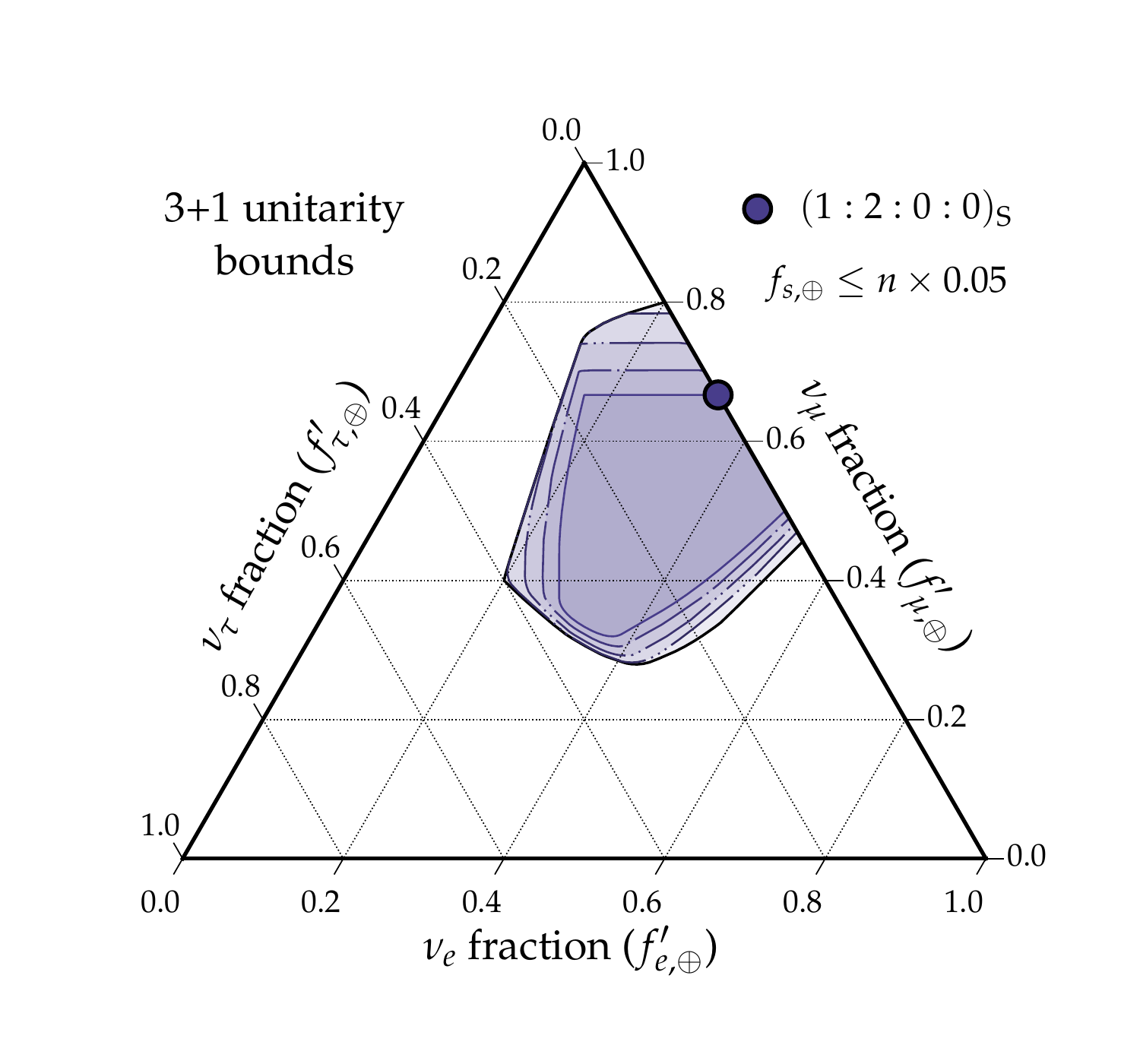}\hspace{0.3cm}\includegraphics[width=0.47\linewidth,viewport=45 30 400 365,clip=true]{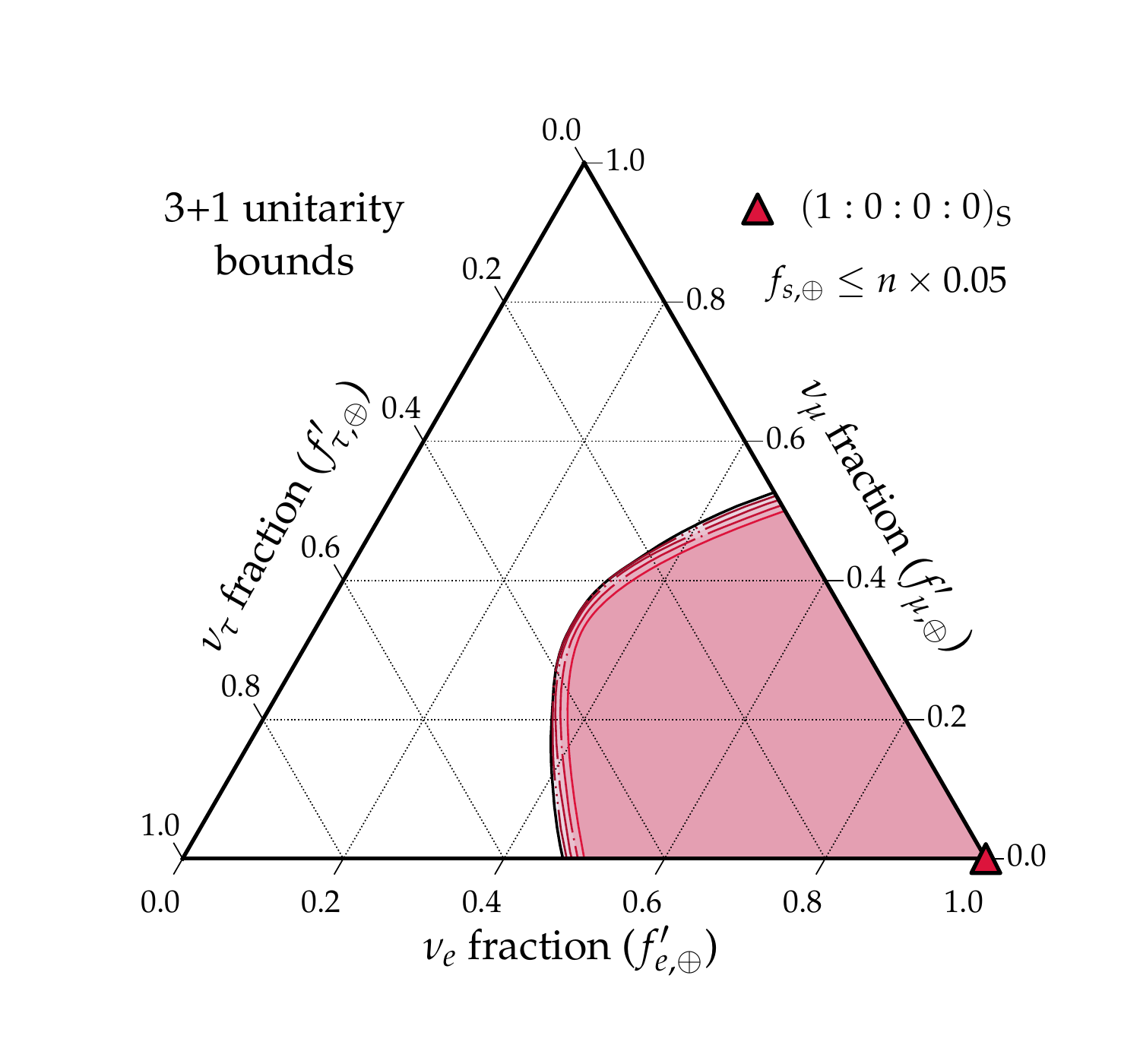}
\caption[]{Flavor boundaries for the source flavor composition $($1$\,:\,$2$\,:\,$0$:\,$0$)_{\rm S}$ (left) and $($1$\,:\,$0$\,:\,$0$:\,$0$)_{\rm S}$ (right) while imposing limits on the sterile neutrino fraction at Earth, $f_{s,\oplus}$. The boundaries are shown for $f_{s,\oplus}\leq n\times0.05$ with integer $0\leq n\leq3$ indicated by the number of consecutive ``dots'' in the boundary lines and increasingly darker line colors. The boundaries quickly saturate the maximal boundary derived by allowing the full range of $f_{s,\oplus}$ accessible with each choice of source flavor composition, indicated by the outer black solid line.}\label{fig4}
\end{figure*}

Finally, Fig.~\ref{fig4} shows the effect on the flavor boundary of imposing limits on the sterile neutrino fraction at Earth, $f_{s,\oplus}$, which indirectly explores the effect of varying the size of the active-sterile mixing elements $\vert U_{\alpha 4} \rvert$. In principle, a strong active-sterile mixing would be inconsistent with the observation of cosmic neutrinos in the first place, since most the flux would have transitioned into sterile neutrinos. For the source flavor composition of $($1$\,:\,$2$\,:\,$0$:\,$0$)_{\rm S}$ (left plot) and $($1$\,:\,$0$\,:\,$0$:\,$0$)_{\rm S}$ (right plot) used in this figure, the contours of increasing $f_{s,\oplus}$ in steps of 5\% grow rapidly to match the boundary derived when allowing the full range of $f_{s,\oplus}$ accessible with each choice of source flavor composition (solid black lines). Therefore, even small overall contributions of sterile neutrinos can leave visible effects on the active flavor fraction.

\section{Conclusions}\label{sec4}

Neutrino mixing has been thoroughly tested at MeV--GeV energies, but remains largely untested at higher energies. The TeV--PeV astrophysical neutrinos discovered by IceCube allow us to test neutrino mixing in a new energy regime. Using them, we may probe a host of models of nonstandard mixing that affect flavor oscillations over cosmological-scale distances. To do this, neutrino telescopes measure the flavor composition of the high-energy astrophysical neutrino flux at Earth. 

We have focused on a large class of theoretically and experimentally motivated models that contain an extra, sterile neutrino ($\nu_s$) that mixes with the three active ones ($\nu_e$, $\nu_\mu$, $\nu_\tau$), while preserving the unitarity of the four-neutrino system. These ``3+1'' models preserve the total number of neutrinos, but redistribute them among the four flavors en route to Earth, modifying the flavor composition compared to the scenario where there is only mixing between the three active flavors. Based solely on the unitarity of the mixing, we have analytically constructed the boundaries that enclose the region of allowed flavor composition at Earth. Via numerical simulations, we have validated that our boundaries tightly enclose the collection of flavor compositions computed by randomly sampling the mixing parameters.

The key advantage of our procedure is that our flavor boundaries are derived analytically, and do not require sampling over the unknown values of the parameters that control active-sterile mixing.  This allows us to test active-sterile mixing at high energies unbiased by the choice of parameter sampling strategy and on its own, independently of existing constraints coming from MeV--GeV experiments. Our boundaries can be used as informed priors in searches for new physics in neutrino telescopes. This extends our earlier study, where we focused on three-flavor unitary mixing~\cite{Ahlers:2018yom}. For convenience, we make the three-flavor and 3+1 boundaries for three popular benchmark scenarios of flavor composition emitted by the sources --- $($1$\,:\,$2$\,:\,$0$:\,$0$)_{\rm S}$, $($0$\,:\,$1$\,:\,$0$:\,$0$)_{\rm S}$, and $($1$\,:\,$0$\,:\,$0$:\,$0$)_{\rm S}$ --- (see Fig.~\ref{fig1}) as well as one nonstandard source scenario following $((1-f_s)/3:(2-2f_s)/3:0:f_s)_{\rm S}$ (see the left panel in Fig.~\ref{fig3}) available as ancillary files.

Our results show that, if no $\nu_s$ are produced at the astrophysical sources, as expected from standard neutrino production processes, the unitarity boundary of allowed flavor composition at Earth in the 3+1 scenario is only marginally extended compared to the three-flavor mixing. This is a surprising result, as the 3+1 scenario has 5 additional degrees of freedom. Standard astrophysical sources with $(f_e:1-f_e:0:0)_{\rm S}$ are still strongly limited in the 3+1 scenario to $f_{\tau, \oplus}\lesssim 0.53$. This can be tested by future IceCube data with higher sensitivity to individual flavors\cite{Song:2020nfh}. On the other hand, we find that including $\nu_s$ production at the source via nonstandard processes allows us to increase the boundaries as long as $f_{s,{\rm S}}\gtrsim 0.5$.

\acknowledgments
The authors would like to thank Poul Henrik Damgaard for support. M.A.~and M.B.~acknowledge support from \textsc{Villum Fonden} under projects no.~18994 and no.~13164, respectively.

\appendix

\setlength{\tabcolsep}{4pt}
\begin{table*}[t!]\centering
\begin{minipage}[t]{\linewidth}
\scriptsize
\begin{tabular}{ccccccc}
\multicolumn{7}{c}{Representations of local extremal points $\widehat{\bf Q}$}\\[0.2cm]
$\begin{pmatrix}
 \,\frac{1}{4}\, &  \frac{1}{4} & \frac{1}{4} & \,\frac{1}{4}\, \\
 \frac{1}{4} & \frac{1}{4} & \frac{1}{4} &\frac{1}{4} \\
 \frac{1}{4} & \frac{1}{4} & \frac{1}{4} & \frac{1}{4} \\
 \frac{1}{4} & \frac{1}{4} & \frac{1}{4} &\frac{1}{4} \\
 \end{pmatrix}$
&
$\begin{pmatrix}
 \mathbf{0}   & \frac{1}{3} & \frac{1}{3} & \frac{1}{3} \\
 \times & \times & \times & \times \\
 \times & \times & \times & \times \\
 \times & \times & \times & \times \\
\end{pmatrix}$
&
$\begin{pmatrix}
 \mathbf{0}   & \mathbf{0}   & \frac{1}{2} & \frac{1}{2} \\
 \times & \times & \times & \times \\
 \times & \times & \times & \times \\
 \times & \times & \times & \times \\
\end{pmatrix}$
&
$\begin{pmatrix}
 \mathbf{0}   & \frac{1}{3} & \frac{1}{3} & \frac{1}{3} \\
 \mathbf{0}   & \frac{1}{3} & \frac{1}{3} & \frac{1}{3} \\
 \times & \times & \times & \times \\
 \times & \times & \times & \times \\
\end{pmatrix}$
&
$\begin{pmatrix}
 \mathbf{0} & \times & \times & \times \\
 \times & \mathbf{0} & \times & \times \\
 \times & \times & \times & \times \\
 \times & \times & \times & \times \\
\end{pmatrix}$
&
$\begin{pmatrix}
 \mathbf{0} & \mathbf{0} & \frac{1}{2} & \frac{1}{2} \\
 \mathbf{0} & \times & \times & \times \\
 \times & \times & \times & \times \\
 \times & \times & \times & \times \\
\end{pmatrix}$
&
$\begin{pmatrix}
 \mathbf{0} & \mathbf{0} & \times & \times \\
 \times & \times & \mathbf{0} & \times \\
 \times & \times & \times & \times \\
 \times & \times & \times & \times \\
\end{pmatrix}_{\!\!\!\Box}$
\\[1cm]
$\begin{pmatrix}
 \mathbf{0} & \times & \times & \times \\
 \mathbf{0} & \times & \times & \times \\
 \times & \mathbf{0} & \times & \times \\
 \times & \times & \times & \times \\
\end{pmatrix}_{\!\!\!\Box}$
&
$\begin{pmatrix}
 \mathbf{0} & \times & \times & \times \\
 \times & \mathbf{0} & \times & \times \\
 \times & \times & \mathbf{0} & \times \\
 \times & \times & \times & \times \\
\end{pmatrix}$
&
$\begin{pmatrix}
 \mathbf{0} & \mathbf{0} & \times & \times \\
 \mathbf{0} & \times & \mathbf{0} & \times \\
 \times & \times & \times & \times \\
 \times & \times & \times & \times \\
\end{pmatrix}_{\!\!\!\Box}$
&
$\begin{pmatrix}
 \mathbf{0} & \mathbf{0} & \frac{1}{2} & \frac{1}{2} \\
 \mathbf{0} & \times & \times & \times \\
 \times & \mathbf{0} & \times & \times \\
 \times & \times & \times & \times \\
\end{pmatrix}_{\!\!\!\Box}$
&
$\begin{pmatrix}
 \mathbf{0} & \mathbf{0} & \times & \times \\
 \mathbf{0} & \times & \times & \times \\
 \times & \times & \mathbf{0} & \times \\
 \times & \times & \times & \times \\
\end{pmatrix}_{\!\!\!\Box}$
&
$\begin{pmatrix}
 \mathbf{0} & \mathbf{0} & \frac{1}{2} & \frac{1}{2} \\
 \frac{1}{2} & \frac{1}{2} & \mathbf{0} & \mathbf{0} \\
 \times & \times & \times & \times \\
 \times & \times & \times & \times \\
\end{pmatrix}$
&
$\begin{pmatrix}
 \mathbf{0} & \mathbf{0} & \times & \times \\
 \times & \times & \mathbf{0} & \times \\
 \times & \times & \mathbf{0} & \times \\
 \times & \times & \times & \times \\
\end{pmatrix}_{\!\!\!\Box}$
\\[1cm]
$\begin{pmatrix}
 \mathbf{0} & \mathbf{0} & \times & \times \\
 \times & \times & \mathbf{0} & \times \\
 \times & \times & \times & \mathbf{0} \\
 \times & \times & \times & \times \\
\end{pmatrix}_{\!\!\!\Box}$
&
$\begin{pmatrix}
 \mathbf{0} & \times & \times & \times \\
 \mathbf{0} & \times & \times & \times \\
 \times & \mathbf{0} & \times & \times \\
 \times & \mathbf{0} & \times & \times \\
\end{pmatrix}$
&
$\begin{pmatrix}
 \mathbf{0} & \times & \times & \times \\
 \mathbf{0} & \times & \times & \times \\
 \times & \mathbf{0} & \times & \times \\
 \times & \times & \mathbf{0} & \times \\
\end{pmatrix}_{\!\!\!\Box}$
&
$\begin{pmatrix}
 \mathbf{0} & \times & \times & \times \\
 \times & \mathbf{0} & \times & \times \\
 \times & \times & \mathbf{0} & \times \\
 \times & \times & \times & \mathbf{0} \\
\end{pmatrix}$
&
$\begin{pmatrix}
 \mathbf{0} & \mathbf{0} & \times & \times \\
 \mathbf{0} & \times & \mathbf{0} & \times \\
 \times & \mathbf{0} & \times & \times \\
 \times & \times & \times & \times \\
\end{pmatrix}_{\!\!\!\Box}$
&
$\begin{pmatrix}
 \mathbf{0} & \mathbf{0} & \times & \times \\
 \mathbf{0} & \times & \mathbf{0} & \times \\
 \times & \times & \times & \mathbf{0} \\
 \times & \times & \times & \times \\
\end{pmatrix}_{\!\!\!\Box}$
&
$\begin{pmatrix}
 \mathbf{0} & \mathbf{0} & \times & \times \\
 \mathbf{0} & \times & \times & \times \\
 \times & \mathbf{0} & \times & \times \\
 \times & \times & \mathbf{0} & \times \\
\end{pmatrix}_{\!\!\!\Box}$
\\[1cm]
$\begin{pmatrix}
 \mathbf{0} & \mathbf{0} & \frac{1}{2} & \frac{1}{2} \\
 \mathbf{0} & \times & \times & \times \\
 \times & \times & \mathbf{0} & \mathbf{0} \\
 \times & \times & \times & \times \\
\end{pmatrix}_{\!\!\!\Box}$
&
$\begin{pmatrix}
 \mathbf{0} & \mathbf{0} & \times & \times \\
 \mathbf{0} & \times & \times & \times \\
 \times & \times & \mathbf{0} & \times \\
 \times & \times & \mathbf{0} & \times \\
\end{pmatrix}_{\!\!\!\Box}$
&
$\begin{pmatrix}
 \mathbf{0} & \mathbf{0} & \times & \times \\
 \mathbf{0} & \times & \times & \times \\
 \times & \times & \mathbf{0} & \times \\
 \times & \times & \times & \mathbf{0} \\
\end{pmatrix}_{\!\!\!\Box}$
&
$\begin{pmatrix}
 \mathbf{0} & \mathbf{0} & \times & \times \\
 \times & \times & \mathbf{0} & \times \\
 \times & \times & \mathbf{0} & \times \\
 \times & \times & \times & \mathbf{0} \\
\end{pmatrix}_{\!\!\!\Box}$
&
$\begin{pmatrix}
 \,\mathbf{0}\, & \mathbf{0} & \frac{1}{2} & \,\frac{1}{2}\, \\
 \mathbf{0} & \mathbf{0} & \frac{1}{2} & \frac{1}{2} \\
 \frac{1}{2} & \frac{1}{2} & \mathbf{0} & \mathbf{0} \\
 \frac{1}{2} & \frac{1}{2} & 0 & 0 \\
\end{pmatrix}$
&
$\begin{pmatrix}
 \mathbf{0} & \mathbf{0} & \times & \times \\
 \mathbf{0} & \times & \mathbf{0} & \times \\
 \times & \mathbf{0} & \mathbf{0} & \times \\
 \times & \times & \times & \times \\
\end{pmatrix}_{\!\!\!\Box}$
&
$\begin{pmatrix}
 \mathbf{0} & \mathbf{0} & \times & \times \\
 \mathbf{0} & \times & \mathbf{0} & \times \\
 \times & \mathbf{0} & \times & \mathbf{0} \\
 \times & \times & \times & \times \\
\end{pmatrix}_{\!\!\!\Box}$
\\[1cm]
$\begin{pmatrix}
 \mathbf{0} & \mathbf{0} & \times & \times \\
 \mathbf{0} & \times & \mathbf{0} & \times \\
 \times & \mathbf{0} & \times & \times \\
 \times & \times & \mathbf{0} & \times \\
\end{pmatrix}_{\!\!\!\Box}$
&
$\begin{pmatrix}
 \mathbf{0} & \mathbf{0} & \times & \times \\
 \mathbf{0} & \times & \mathbf{0} & \times \\
 \times & \mathbf{0} & \times & \times \\
 \times & \times & \times & \mathbf{0} \\
\end{pmatrix}_{\!\!\!\Box}$
&
$\begin{pmatrix}
 \mathbf{0} & \mathbf{0} & \times & \times \\
 \mathbf{0} & \times & \mathbf{0} & \times \\
 \times & \times & \times & \mathbf{0} \\
 \times & \times & \times & \mathbf{0} \\
\end{pmatrix}_{\!\!\!\Box}$
&
$\begin{pmatrix}
 \mathbf{0} & \mathbf{0} & \frac{1}{2} & \frac{1}{2} \\
 \mathbf{0} & \times & \times & \times \\
 \times & \mathbf{0} & \times & \times \\
 \times & \times & \mathbf{0} & \mathbf{0} \\
\end{pmatrix}_{\!\!\!\Box}$
&
$\begin{pmatrix}
 \mathbf{0} & \mathbf{0} & \times & \times \\
 \mathbf{0} & \times & \times & \times \\
 \times & \times & \mathbf{0} & \mathbf{0} \\
 \times & \times & \mathbf{0} & \times \\
\end{pmatrix}_{\!\!\!\Box}$
&
$\begin{pmatrix}
 \mathbf{0} & \mathbf{0} & \times & \times \\
 \mathbf{0} & \times & \mathbf{0} & \times \\
 \times & \mathbf{0} & \mathbf{0} & \times \\
 \times & \times & \times & \mathbf{0} \\
\end{pmatrix}_{\!\!\!\Box}$
&
$\begin{pmatrix}
 \mathbf{0} & \mathbf{0} & \times & \times \\
 \mathbf{0} & \times & \mathbf{0} & \times \\
 \times & \mathbf{0} & \times & \mathbf{0} \\
 \times & \times & \mathbf{0} & \times \\
\end{pmatrix}_{\!\!\!\Box}$
\\[1cm]
$\begin{pmatrix}
 \,\mathbf{0} \, & \mathbf{0} & \frac{1}{2} &  \,\frac{1}{2} \, \\
 \mathbf{0} & \frac{1}{2} & \mathbf{0} & \frac{1}{2} \\
 \frac{1}{2} & \mathbf{0} & \frac{1}{2} & \mathbf{0} \\
 \frac{1}{2} & \frac{1}{2} & \mathbf{0} & \mathbf{0} \\
\end{pmatrix}_{\!\!\!\Box}$
&
$\begin{pmatrix}
 \,1 \, & \mathbf{0}   & \mathbf{0}   &  \,\mathbf{0} \,   \\
 \mathbf{0}   & \frac{1}{3} & \frac{1}{3} & \frac{1}{3} \\
 \mathbf{0}   & \frac{1}{3} & \frac{1}{3} & \frac{1}{3} \\
 0 & \frac{1}{3} & \frac{1}{3} & \frac{1}{3}  \\
\end{pmatrix}$
&
$\begin{pmatrix}
 1& \mathbf{0}   & \mathbf{0}   & \mathbf{0}   \\
 \mathbf{0}   & \mathbf{0}   & \frac{1}{2}  & \frac{1}{2} \\
 \mathbf{0}   & \times & \times & \times \\
 0 & \times & \times & \times \\
\end{pmatrix}$
&
$\begin{pmatrix}
 1& \mathbf{0}   & \mathbf{0}   & \mathbf{0}   \\
 \mathbf{0}   & \mathbf{0}   & \times  & \times \\
 \mathbf{0}   & \times &  \mathbf{0} & \times \\
 0 & \times & \times & \times \\
\end{pmatrix}_{\!\!\!\triangle}$
&
$\begin{pmatrix}
  \,1 \,& \mathbf{0}   & \mathbf{0}   &  \,\mathbf{0} \,   \\
 \mathbf{0}   & \mathbf{0}   & \frac{1}{2}  & \frac{1}{2} \\
 \mathbf{0}   & \frac{1}{2}   & \mathbf{0} & \frac{1}{2}   \\
 0 & \frac{1}{2} & \frac{1}{2} &\mathbf{0} \\
\end{pmatrix}_{\!\!\!\triangle}$
&
$\begin{pmatrix}
 \,1 \, & \mathbf{0}   & \mathbf{0}   &  \,\mathbf{0} \,   \\
 \mathbf{0}   & 1 & \mathbf{0}   & \mathbf{0}   \\
 \mathbf{0}   & \mathbf{0}   & \frac{1}{2}  & \frac{1}{2} \\
 0 & 0 & \frac{1}{2}  & \frac{1}{2} \\
\end{pmatrix}$
&
$\begin{pmatrix}
  \,1 \, & \mathbf{0}   & \mathbf{0}   &  \,\mathbf{0} \,   \\
 \mathbf{0}   & 1 & \mathbf{0}   & \mathbf{0}   \\
 \mathbf{0}   & \mathbf{0}   & 1 & \mathbf{0}   \\
 0 & 0 & 0 & 1\\
\end{pmatrix}$\\
\end{tabular}
\end{minipage}
\caption[]{The structure of local extremal points $\widehat{\bf Q}$ in the 3+1 mixing scenario representing equivalence classes of solutions. The boldface entries indicate the boundary conditions $Q_{\alpha_ti_t}=0$ fixed by a set of Lagrange multipliers $\lambda_t$ (see Eq.~(\ref{eq:Lagrange})) in the solution (\ref{eq:finalQ}). All entries that evaluate to constant values are also indicated; the entries indicated by symbols ``$\times$'' are functions of the parameters $u$, $v$, $w$, $x$, $y$, and $z$. All solutions with the symbol ``$\Box$'' in the bottom-right corner violate the quadrilateral inequality, Eq.~(\ref{eq:bound2}), in at least one pair of columns or rows. The last six matrices in the bottom row correspond to solutions inherited from the three-flavor mixing scenario discussed in Ref.~\cite{Ahlers:2018yom}. Two of these matrices, indicated by the symbol ``$\triangle$'', violate the three-flavor triangle inequality and can be neglected, resulting in 40 remaining solutions.}\label{tab1}
\end{table*}

\section{Unitarity Bounds}\label{appI}

For the derivation of the boundary function of Eq.~(\ref{eq:bound}) we follow the procedure outlined in Ref.~\cite{Ahlers:2018yom}. The oscillation-averaged neutrino flavor-transition matrix can be written as the matrix product
\begin{equation}\label{eq:PQQ}
{\bf P}={\bf Q}{\bf Q}^T\,,
\end{equation}
where ${Q}_{\alpha i} \equiv |U_{\alpha i}|^2$. The matrix elements of ${\bf Q}$ are subject to the unitarity condition ${\bf U}^\dagger{\bf U} = {\boldsymbol 1}$. This imposes the normalization condition 
\begin{equation}\label{eq:norm}
\sum_\alpha{Q}_{\alpha i}=\sum_i{Q}_{\alpha i}=1\,,
\end{equation}
and the boundary condition
\begin{equation}\label{eq:bound1}
0\leq{Q}_{\alpha i}\leq1\,.
\end{equation}
In addition, from the off-diagonal elements of the unitarity condition ${\bf U}^\dagger{\bf U} = {\boldsymbol 1}$, the elements of ${\bf Q}$ are subject to quadrilateral inequalities that can be summarized by the six conditions ($\alpha<\beta$)
\begin{equation}\label{eq:bound2}
\mathcal{A}(\sqrt{Q_{\alpha 1}Q_{\beta 1}},\sqrt{Q_{\alpha 2}Q_{\beta 2}},\sqrt{Q_{\alpha 3}Q_{\beta 3}},\sqrt{Q_{\alpha 4}Q_{\beta 4}})\geq0\,,
\end{equation}
where the function $\mathcal{A}$ is defined as 
\begin{equation}\label{eq:T}
\mathcal{A}(a,b,c,d) \equiv (a+b+c-d)(b+c+d-a)\times(c+d+a-b)(d+a+b-c)\,,
\end{equation}
and is proportional to the square of the {\it maximum} area of the quadrilateral with sides $a$, $b$, $c$, and $d$. 

The most stringent bound $B(u,v,w,x,y,z)$ in Eq.~(\ref{eq:bound}) corresponds to the global maximum of the function
\begin{equation}\label{eq:G}
G({\bf Q};{\bf c}) =\frac{1}{2}\text{Tr}\left({\bf c}\,{\bf Q}{\bf Q}^T\right)\,,
\end{equation}
for a fixed coefficient matrix
\begin{equation}\label{eq:epsilon}
{\bf c} = \begin{pmatrix}0&z&y&u\\z&0&x&v\\y&x&0&w\\u&v&w&0\end{pmatrix}\,,
\end{equation}
and subject to the normalization (\ref{eq:norm}) and boundary conditions (\ref{eq:bound1}) and (\ref{eq:bound2}). We follow closely the procedure outlined in Ref.~\cite{Ahlers:2018yom} by first identifying all local extrema $\widehat{\bf Q}$ of Eq.~(\ref{eq:G}) that saturate the (technically) simple boundaries (\ref{eq:bound1}), {\it i.e.}, either $Q_{\alpha i} = 0$ or 1, and then finding the global maximum among them. In the three-flavor scenario discussed in Ref.~\cite{Ahlers:2018yom} it could be shown that this procedure always yields accurately the global maximum of $G$. However, in the 3+1 scenario, as we have shown in the main text, this condition is, in general, not satisfied. Nevertheless, our procedure still allows us to derive nontrivial flavor boundaries $B(u,v,w,x,y,z)$ that enclose all possible flavor compositions at Earth, for a given flavor composition at the source.

As in the three-flavor scenario~\cite{Ahlers:2018yom}, we make use of the fact that the set of local extrema of Eq.~(\ref{eq:G}) is invariant under the transformation
\begin{equation}\label{eq:invariance}
\widehat{\bf Q} \to \widehat{\bf Q}' \equiv {\bf F}\widehat{\bf Q}{\bf M}^T\qquad {\bf c} \to {\bf c}' \equiv {\bf F}{\bf c}{\bf F}^T\,,
\end{equation}
where ${\bf F}$ and ${\bf M}$ are two permutation matrices of the flavor and mass indices, respectively. In other words, the solutions are invariant under the exchange of entries of two arbitrary columns of $\widehat{\bf Q}$ and the corresponding reordering of entries in ${\bf c}$. Therefore, in the following, we derive separate equivalence classes of solutions Eq.~(\ref{eq:G}): within each class, solutions are related by the transformations (\ref{eq:invariance}).  The full list of candidate extrema of Eq.~(\ref{eq:G}) can then be recovered by applying these transformations within each class.

We systematically search for all local extrema of Eq.~(\ref{eq:G}) that saturate the conditions (\ref{eq:bound1}) by the method of Lagrange multipliers. The solutions are of the form 
\begin{equation}\label{eq:finalQ}
\widehat{Q}_{\alpha i} = \frac{1}{4} - \frac{1}{4}\sum_{t=1}^n\lambda_t A_{\alpha \alpha_t}B_{i i_t}\,,
\end{equation}
where $\lambda_t$ are $n$ Lagrange multipliers for the boundary condition $Q_{\alpha_ti_t}=0$ and we define the matrices
\begin{align}
A_{\alpha\beta} &\equiv (c^{-1})_{\alpha\beta} - \frac{\sum\limits_{\mu}(c^{-1})_{\mu\alpha} \sum\limits_{\nu}(c^{-1})_{\nu\beta}}{\sum\limits_{\mu\nu}(c^{-1})_{\mu\nu}}\,,\\
B_{ij} &\equiv \delta_{ij} - \frac{1}{4}\,.
\end{align}
Note that the normalization condition (\ref{eq:norm}) follows from the identities $\sum_iB_{ij}=\sum_\alpha A_{\alpha\beta}=0$. The value of the Lagrange multipliers follows from the conditions $Q_{\alpha_ti_t}=0$ and expression~(\ref{eq:finalQ}), {\it i.e.},
\begin{equation}\label{eq:Lagrange}
\lambda_t = \frac{1}{\text{det}({\bf C})}\sum_{s=1}^n(-1)^{s+t}\text{det}({\bf C}_{st})\,,
\end{equation}
where $C_{st}\equiv A_{\alpha_s\alpha_t}B_{i_s i_t}$ and ${\bf C}_{ij}$ denotes the remaining sub-matrix after removal of the $i$-th row and $j$-th column. 

We identify 42 equivalence classes of extremal points $\widehat{\bf Q}$ that are summarized in Table.~\ref{tab1}. The boldface matrix element in each class indicate the index pair $(\alpha_t,i_t)$ that is fixed via Lagrange multipliers $\lambda_t$ in Eq.~(\ref{eq:finalQ}).  We also highlight those entries that evaluate to constant values. All other entries, marked by the symbol ``$\times$'', are functions of the coefficients $u$, $v$, $w$, $x$, $y$, and $z$ from Eq.~(\ref{eq:epsilon}). Each solution $\widehat{\bf Q}_i$ listed in Table~\ref{tab1} corresponds to a local extremum $G_i({\bf c}) \equiv G(\widehat{\bf Q}_i;{\bf c})$. The full set of local extrema in each class can then be reconstructed by applying the transformations~(\ref{eq:invariance}) to the entries of the second column, {\it i.e.},
\begin{equation}
\mathcal{S}_i \equiv \lbrace G_i({\bf c}' = {\bf F}{\bf c}{\bf F}^T)\,\big|\,{\bf F}\in S_4\rbrace\,,
\end{equation}
where ${\bf F}$ are matrices of the permutation group $S_4$.

While we have presented the procedure for the 3+1 scenario, it can be easily extended to the $3+n$ scenario, with $n > 1$ and an extended set of local equivalence classes.
 
\section{Active Flavor Projection}\label{appII}

Unitarity in the 3+1 flavor scenario sets a bound on the space of flavor shifts $(\Delta f_e,\Delta f_\mu,\Delta f_s)$ from a given source flavor composition. We can visualize these bounds in a ternary diagram of active flavor fractions $(f_e^\prime, f_\mu^\prime,f_\tau^\prime)$ with $f_{\alpha}' \equiv f_{\alpha}/(1-f_s)$ and $f_e^\prime+ f_\mu^\prime+f_\tau^\prime=1$. To do this, we parametrize the two-dimensional shift in the space of active flavors via a direction $\alpha$ and range $\ell$,
\begin{align}\label{eq:eshift}
\Delta f'_e &\equiv f'_{e,\oplus} - f'_{e,{\rm S}} \equiv \ell\cos\alpha\,,\\\label{eq:mushift}
\Delta f'_\mu &\equiv f'_{\mu,\oplus} - f'_{\mu,{\rm S}} \equiv \ell\sin\alpha\,.
\end{align}
Our goal is to identify the maximum range $\ell$ of an active flavor shift in the direction $\alpha$. The resulting parametric solution $\ell_{\rm max}(\alpha)$ then corresponds to the projection of the 3+1 unitarity bound onto the space of active flavor fractions via Eqs.~(\ref{eq:eshift}) and (\ref{eq:mushift}).

Using Eqs.~(\ref{eq:eshift}) and (\ref{eq:mushift}), the three-dimensional flavor shift is 
\begin{align}
\Delta f_e &= \ell (1-f_{s,\oplus})\cos\alpha - \Delta f_s f'_{e,{\rm S}}\,,\\
\Delta f_\mu &= \ell (1-f_{s,\oplus})\sin\alpha - \Delta f_s f'_{\mu,{\rm S}}\,,\\
\Delta f_s &=f_{s,\oplus}-f_{s,{\rm S}}\,.
\end{align}
We now look at the three-dimensional unitarity boundary $\widehat{\bf n} \cdot (\Delta f_e,\Delta f_\mu,\Delta f_s) = B(\widehat{\bf n})$, where $\widehat{\bf n}$ is a unit vector corresponding to the normal of the three-dimensional boundary surface.  For a fixed sterile fraction $f_{s,\oplus}$ and $\widehat{\bf n}$, the maximum flavor shift $\ell$ in the direction $\alpha$ is
\begin{equation}\label{eq:ellalpha}
\ell(\alpha,f_{s,\oplus},\widehat{\bf n}) = \frac{B(\widehat{\bf n}) + (f_{s,\oplus}-f_{s,{\rm S}})(\widehat{n}_ef'_{e,{\rm S}}+\widehat{n}_\mu f'_{\mu,{\rm S}}-\widehat{n}_s)}{(1-f_{s,\oplus})(\widehat{n}_e\cos\alpha +\widehat{n}_\mu\sin\alpha )}\,.
\end{equation}
The two-dimensional boundary in the direction $\alpha$ is determined from the set of all solutions (\ref{eq:ellalpha}) by first finding the minimum with respect to all boundaries $B(\widehat{\bf n})$ for a fixed sterile fraction $f_{s,\oplus}$,
\begin{equation}
\ell_{\rm min}(\alpha,f_{s,\oplus}) = \min_{\widehat{\bf n}}\Big(\Big\lbrace{\ell(\alpha,f_{s,\oplus},\widehat{\bf n})\Big||\Delta\alpha| < \frac{\pi}{2}\Big\rbrace}\Big)\,,
\end{equation}
where $\cos\Delta\alpha = \widehat{n}_e\cos\alpha +\widehat{n}_\mu\sin\alpha$, and then by finding the maximum with respect to all accessible sterile fractions, {\it i.e.},
\begin{equation}\label{eq:ell}
\ell_{\rm max}(\alpha) = \max\limits_{f_{s,\oplus}}\Big(\Big\lbrace\ell_{\rm min}(\alpha,f_{s,\oplus})\Big\rbrace\Big)\,.
\end{equation}
By varying $\alpha$, we plot the two-dimensional boundaries in Figs.~\ref{fig1}--\ref{fig3} in the main text.

\bibliographystyle{utphys_mod}

\bibliography{references}
\end{document}